# GapPredict – A Language Model for Resolving Gaps in Draft Genome Assemblies

Eric Chen, Justin Chu, Jessica Zhang, René L. Warren, Inanc Birol

**Abstract**— Short-read DNA sequencing instruments can yield over $10^{12}$ bases per run, typically composed of reads 150 bases long. Despite this high throughput, *de novo* assembly algorithms have difficulty reconstructing contiguous genome sequences using short reads due to both repetitive and difficult-to-sequence regions in these genomes. Some of the short read assembly challenges are mitigated by scaffolding assembled sequences using paired-end reads. However, unresolved sequences in these scaffolds appear as "gaps". Here, we introduce GapPredict – a tool that uses a character-level language model to predict unresolved nucleotides in scaffold gaps. We benchmarked GapPredict against the state-of-the-art gap-filling tool Sealer, and observed that the former can fill 65.6% of the sampled gaps that were left unfilled by the latter, demonstrating the practical utility of deep learning approaches to the gap-filling problem in genome sequence assembly.

✦

## 1 INTRODUCTION

THE emergence of next-generation, high-throughput genome sequencing technologies has revolutionized life sciences. In next-generation sequencing, state-of-the-art instruments read genomes at high-depth, but yield relatively short (Illumina, ~150 bp) individual sequence segments "reads", providing unprecedented volumes of sequencing data [1]. In order to reconstruct the input genomes, these short sequencing reads need to be assembled together and unbiased genome assemblies are performed *de novo* – without the use of a reference genome [2]. *De novo* genome assembly remains an open problem, with leading algorithms in the field usually yielding partial and incomplete genome sequences [3].

*De novo* assembly algorithms identify partial and unambiguous read-to-read overlaps to merge and extend the reads into contiguous sequences, or *contigs*. Many tools, such as ABySS [4] and SPAdes [5], are designed to perform *de novo* assembly on a set of short paired-end reads. Typically, paired-end reads are short sequences (< 250 bp) generated from the two ends of a DNA fragment several hundred bases in length.

For complex genomes such as the human genome, although *de novo* assemblers have successfully produced draft genome assemblies, they are incomplete [4]. Often, these *de novo* assemblies contain many gaps, which are regions of unknown nucleotide sequence [3], [4]. *De novo* assemblers generate gaps during the process of scaffolding, where flanking sequences are inferred to follow each other. Still, precise sequence content between the flanking regions may remain undetermined. Gaps are caused by factors such as local depressions in the read coverage depth resulting in missed read-to-read overlaps and are often also caused by the inability of short reads to resolve repetitive sequences in these regions [6]. Filling gaps in *de novo* assemblies improves the quality of draft genomes, which has implications for downstream analyses such as structural variation identification [7], [8], [9], [10] and gene annotation [6].

Gap-filling is a well-studied problem, and there are established tools, such as Sealer [6] and GAPPadder [11], that provide solutions, albeit with varied performance, which are influenced by many of the intrinsic factors noted above and limitations associated with their implementation. For example, Sealer is reported to fill 50.8% of about 240,000 gaps in a human genome assembly, and 13.8% of about 3 million gaps in a white spruce genome assembly draft [6]. Gap-filling tools tend to use greedy algorithms to solve the gap-filling problem [6], [11]. Greedy algorithms are limited by their rigid heuristics, and may be unable to fully exploit all the information contained in short read sequencing data. As a result, these state-of-the-art tools still leave many gaps unfilled during assembly of complex genomes. To better exploit short read data, we investigated a different paradigm to the gap-filling problem, using a deep learning approach.

Deep learning is the use of neural networks – data-driven, tunable functions – to allow computers to extract features from a dataset and make predictions on similar data [12]. Deep learning has already seen successful application in computational biology, especially in sequence classification tasks [13], [14], [15], [16]. However, few applications of deep learning in computational biology seem to exist for sequence prediction. One such application, HELEN [17], uses a single recurrent neural network trained on read-to-assembly alignment summary statistics obtained from MarginPolish [17] to fix base assembly errors in long read assembly drafts.

---
- E. Chen is with the Genome Sciences Centre, BC Cancer, Vancouver, BC V5Z 4S6, Canada. E-mail: ericchen@alumni.ubc.ca.
- J. Chu is with the Genome Sciences Centre, BC Cancer, Vancouver, BC V5Z 4S6, Canada. E-mail: cjustin@bcgsc.ca.
- J. Zhang is with the Genome Sciences Centre, BC Cancer, Vancouver, BC V5Z 4S6, Canada. E-mail: jczhang97@alumni.ubc.ca.
- R.L. Warren is with the Genome Sciences Centre, BC Cancer, Vancouver, BC V5Z 4S6, Canada. E-mail: rwarren@bcgsc.ca.
- I. Birol is with the Genome Sciences Centre, BC Cancer, Vancouver, BC V5Z 4S6, Canada. E-mail: ibirol@bcgsc.ca.

The gap-filling problem can be framed as a sequence prediction problem, as the sequence preceding the gap may provide sufficient context to predict the gap sequence itself. We note that the gap-filling problem typically utilizes large volumes of data to represent the sequence content of gaps - an ideal condition to leverage deep learning approaches [12].

Our objective with this study is to assess the suitability of supervised deep learning algorithms for the gap-filling problem. To explore this possibility, we introduce GapPredict, a character-level language model for filling gaps in draft genome assemblies. Character level language models predict the most likely character from a corpus of characters after receiving a sequence of characters from that corpus as context [18].

In this study, we benchmarked the performance of GapPredict against Sealer [6] and GAPPadder [11] – two scalable heuristic-based gap-filling tools. We observed that GapPredict compares favorably to both Sealer and GAPPadder. This demonstrates that deep learning may be a viable approach for the gap-filling task in genome assembly.

## 2 IMPLEMENTATION

### 2.1 Overview

GapPredict takes two files as input – a FASTA file containing the two sequences flanking a given gap (henceforth referred to as flanks or flanking sequences), and a FASTQ file containing paired-end reads mapping to the flanks of the gap. Note that we imposed no constraints on input flank length, gap length, or read length. The reads and their reverse-complements are used to train a language model. Provided that the gap is shorter than the fragment length of the reads, the reads in the FASTQ file collectively span both the gap flanks and the gap completely. Thus, after training on the reads, GapPredict should have sufficient data to fill the gap in either the forward or reverse-complement direction if given a flank as context.

Following training, GapPredict uses its language model to recursively predict the sequence in a given gap using one of the flank sequences as its initial context. Both the forward and reverse-complement of the gap can be predicted by GapPredict, as the model may predict one direction better than the other (Fig. 1).

### 2.2 Language Model Architecture

We implemented the GapPredict model using the Keras framework (v2.2.4; Chollet F; [*https://github.com/keras-team/keras*]) and Tensorflow [19]. The GapPredict model architecture consists of three sequential layers (Fig. S1). First, each base in the input sequence, represented as a one-hot vector, is encoded as a word vector by an embedding layer. Next, the resulting sequence of word vectors is fed into a long short-term memory (LSTM). An LSTM was chosen as this architecture has been shown to offer good performance on tasks involving long sequences [20], [21], which we consider gap-filling to be. Finally, the LSTM state is fed into a fully connected layer of neurons (a "dense layer" [19]). The output of this layer is a vector of length 4, which is normalized by the softmax function. Each value in this output vector can be interpreted as the probability that the next base is one of the four corresponding deoxyribonucleotides (A, C, G, T). We optimized this model using Adam [22], which is known to be a good out-of-the-box function. Our loss function was categorical cross-entropy, as each iteration of our sequence prediction algorithm is a multi-class classification task.

### 2.3 Language Model Training

The training protocol for the GapPredict language model follows a four-step cycle. At each training iteration, we first randomly sample a batch of reads with replacement (Fig. S2). Next, we randomly choose a length $k$ between $k_{low}$ and $k_{high}$, two hyperparameters, and extract a random $k + 1$-mer from each read in the batch (Fig. S3). We then compute the categorical cross-entropy loss for predicting the $k + 1$-th base, given the first $k$ bases and adjusts the model parameters accordingly (Fig. S4).

At the end of every epoch, to inform early stopping, we compute the validation loss as follows. For a given flank of length $f$, we take the first $x$ bases (for all $x \in [k_{low}, f - 1]$) and compute the categorical cross-entropy loss for predicting the $x + 1$-th base given the first $x$ bases (Fig. S5). In essence, $x$ increases iteratively from $k_{low}$ to $f - 1$. The validation loss is the sum of the loss for predicting the bases of every flank divided by the sum of the flank lengths.

Our validation loss metric measures a model's ability to predict each flank and its reverse-complement. We rationalized that if our model was capable of correctly predicting the next base along each flank and on both DNA strands, then it is likely to have encoded information to predict the gap as well.

### 2.4 Gap Sequence Prediction

After training the model for a given gap, we predict the nucleotide sequence of the associated gap. Each gap is predicted with beam search using both the flanking sequence in the forward direction and the flanking sequence in the reverse-complement direction [23] (Fig. S6).

As described above, each base prediction has an associated probability score for how likely the next base is. Thus, taking the log-sum of these probabilities gives us a metric for how confident our model is for the entire sequence it outputs. Beam search provides a scalable and robust, albeit greedy, method of searching for the output sequence with minimal magnitude of log-sum probability.

**Availability:** Source code is available at
https://github.com/bcgsc/GapPredict/releases/tag/v1.0b.

## 3 METHODS

Refer to Fig. S7 for an overview of our pipeline.

### 3.1 Gap Data Acquisition

We used the *de novo* assembler ABySS (abyss-pe v2.1.5) [4] to assemble the NA12878 human genome (Paired-end 250bp sequencing data downloaded from https://basespace.illumina.com, flow cell H00DDBCXX), using a k-mer length of 144 bp. Then, we ran Sealer (abyss-sealer v2.1.5) [6] on our draft assembly to close its gaps.

From the Sealer output, we randomly selected 900 gaps that it filled (set 1) and 900 gaps that it failed to fill (set 2). In our tests, we required the gap flanks to be represented by 500 bp sequences, have unambiguous alignments in the reference human genome, and represent *bona fide* gaps as assessed with respect to the reference genome. After filtering out false gaps and gaps with shorter or ambiguously aligned flanks, we were left with 434 gaps in set 1 and 416 gaps in set 2.

From each gap, we extracted 500 bp flanks from both sides to construct a FASTA file using a combination of in-house scripts, SAMtools (v1.9) [24], and BEDtools (v2.27.1) [25]. Finally, we used the BioBloomMIMaker utility from BioBloom Tools (v2.3.2) [26] to construct a multi-index Bloom filter for each flank. Next, using BioBloomMICategorizer [26] we built a FASTQ file by selecting any read, along with its mate, that mapped to a gap flank sequence. For each gap, this pair of FASTA and FASTQ files was the input used to run GapPredict.

### 3.2 GapPredict Configurations

In our tests, we initialized our GapPredict models with an embedding vector length of 128 and 512 LSTM units. Our models were trained over at most 1000 epochs with a batch size of 128. Early stopping was employed on validation loss with a patience of 200 epochs. $k_{low}$ was set to 52 bp and $k_{high}$ was set to the length of the shortest read of each training batch.

To predict the sequence of a given gap, our model selected each of the 500 bp flanks as input and predicted the next 750 bases using a beam length of 64 bp. While we chose the prediction and beam lengths arbitrarily, our prediction length was long enough to cover both the estimated gap length and most of the reciprocal flank for every gap. We define the reciprocal flank as the flank on the opposite side of the input flank.

### 3.3 Reference Gap Sequence Acquisition

In order to quantify the performance of the tools we benchmarked – GapPredict, Sealer, and GAPPadder – we compared the sequences they predicted to sequences we extracted from the human genome reference HG38. To build this "ground truth" from HG38, for each gap in our benchmarks, we aligned its flanking sequences to HG38 using BWA-MEM (v0.7.17) [27] and SAMtools [24]. We then used BEDtools bamtobed and BEDtools getfasta [25] to obtain the sequence of both the gap and its flanks in HG38.

### 3.4 GapPredict Output Validation

GapPredict is implemented to always output the next N bases (set at 750 bp by default) from the input flank. Since gaps in our experiment were often shorter than 200 bp, GapPredict output should include the candidate gap sequence, followed by the initial bases of the reciprocal flank.

Using Exonerate (v2.2.0) [28] we aligned GapPredict predictions to the reference gap sequence and 100 bp of the reference reciprocal flank. We evaluated these alignments with four metrics: sequence percent identity, target sequence percent coverage, query sequence percent coverage, and sequence percent correctness, where "query" refers to GapPredict predictions and "target" refers to the reference sequence for alignment.

Of course, in a typical use case, one would not have a reference for filled gaps. Thus, we needed a heuristic for deciding if a gap is likely to be correctly filled ("pass"). We reasoned that an accurately predicted gap sequence is the most likely sequence context to yield an accurate prediction of the reciprocal flank [29]. Since we know the sequence of both gap flanks, we define a prediction to be a "pass" if the first 100 bp of the reciprocal flank sequence aligns to GapPredict's prediction for the gap with a sequence percent correctness over 70%, and a "fail" otherwise.

Using Seaborn (v0.9.0; Waskom M et al.; [https://github.com/mwaskom/seaborn]), we compared target sequence percent correctness against query sequence percent coverage for each gap we predicted. We also determined the probability density for these two variables.

### 3.5 Gap Prediction Validation Metrics

Sequence percent identity is the percentage of matches in the alignment to the total number of aligned positions, including gaps, from a given sequence (definition L2 in [30]).

Target sequence coverage is the quotient between the number of reference sequence bases aligned to the prediction and the total reference sequence length. Query sequence coverage is the quotient between the number of bases from the prediction aligned to the reference sequence and the number of bases from the start of our predicted gap to the end of our predicted gap. Our predicted gap is defined to end at the maximum reference base index among where the gap alignment ends and the reciprocal flank alignment begins. In essence, target sequence coverage provides a measure of how much of the reference sequence is covered by GapPredict's prediction, whereas query sequence coverage provides a measure of how many bases predicted by GapPredict are actually related to the reference sequence.

Lastly, sequence percent correctness is the product of sequence percent identity and sequence percent coverage. This metric was computed to aggregate sequence percent identity and sequence coverage.

### 3.6 Sealer Output Validation

Although Sealer can be run on the whole NA12878 assembly and associated read set, we ran Sealer on each gap from sets 1 and 2 individually using only the reads GapPredict utilized as input for its gap predictions. This was done to confirm that using only reads mapping to the gap (to be consistent with GapPredict's workflow), rather than all reads in the NA12878 dataset, had negligible effect on Sealer's gap-filling performance.

We looked at the target percent correctness of Sealer's output gaps when compared to the reference, as reported by Exonerate [28]. This was done to benchmark the performance of heuristic algorithms. Finally, we compared the target percent correctness of both GapPredict and Sealer for each gap we predicted, and determined the probability density for these two variables.

### 3.7 GAPPadder Output Validation

We ran GAPPadder (base version on https://github.com/simoncchu/GAPPadder commit a359750) [11] using the entire NA12878 assembly and reads, rather than on each gap from sets 1 and 2 individually. This is because GAPPadder searches for discordant reads [11], which would not be present for gaps taken in isolation. After GAPPadder filled all gaps it identified in the draft assembly, we used BWA-MEM [27] to map GAPPadder's output to HG38 reference sequences for gaps in sets 1 and 2. We chose the best alignment when a gap filled by GAPPadder mapped ambiguously to the reference genome.

Similar to Sealer, we looked at the target percent correctness of GAPPadder's output gaps when compared to the reference. We also compared the target percent correctness of GapPredict and GAPPadder's outputs and determined the probability density for these two variables.

### 3.8 QUAST Evaluation

We ran the sequence quality assessment tool QUAST [3] (v5.0.2 -m 0 -t 8) separately on predicted sequence outputs from gaps closed in common between Sealer, GAPPadder, and GapPredict using HG38 gap sequences as a reference.

## 4 RESULTS AND DISCUSSION

### 4.1 GapPredict Output Validation

Because GapPredict makes two predictions per gap, one in the forward and one in the reverse-complement direction, running the tool on 434 gaps in set 1 and 416 gaps in set 2 resulted in 868 and 832 predictions for the two sets, respectively. Of these predictions, 78.7% in set 1 and 65.2% in set 2 were classified as a "pass". In addition, for the 434 gaps in set 1, 87.3% had at least one pass in the prediction pair and 70.0% had two passes. For the 416 gaps in set 2, 78.4% had at least one pass in the prediction pair and 52.2% had two passes. The proportion of passes being lower in set 2 reinforces the notion that these gaps may be more challenging to fill.

For gap predictions classified as a "pass" in both set 1 (Fig. 2a) and set 2 (Fig. 3a), there was a significant number of predictions with high target percent correctness and query percent coverage (top right corner). Gap predictions classified as a "fail" in both set 1 (Fig. 2b) and set 2 (Fig. 3b), on the other hand, formed two types of clusters – clusters of high target percent correctness (right side), and clusters of both low target percent correctness and low query percent coverage (bottom left corner). Set 2 also contained a third cluster at high target percent correctness and low query percent coverage.

Note that Fig. 2b and Fig. 3b both have multi-layered contours at the bottom left corner, despite the scatter plot being highly concentrated at this location. 56.8% of points for Fig. 2b and 54.3% of points for Fig. 3b are located at these corners. Because kernel density estimations provide a probability density for our scatter plot [31], the multi-layered contours reflect the high probability of gaps classified as a "fail" being predicted with target percent correctness and query percent coverage close to 0.

When gap predictions are categorized as a "pass", they tend to have high target percent correctness and high query percent coverage. This demonstrates that models that manage to predict the reciprocal flank are also likely to predict the gap itself. Thus, our heuristic for good predictions is valuable for identifying low quality gap predictions. However, because it is possible that our models predict the gap correctly but the reciprocal flank incorrectly, we may miss some high-quality gap predictions. The clusters of "failed" predictions with high target percent correctness and high query percent coverage in Fig. 2b and 3b illustrate this.

### 4.2 Performance Against Other Tools

Sealer's ability to fill gaps did not particularly change when Sealer was executed on gaps using only reads mapping to the gap and its flanks, instead of all reads in the NA12878 dataset (Fig. 4). Rather than filling all 434 gaps from set 1 and none of the 416 gaps from set 2, as with the latter approach, Sealer filled 430 gaps (99.1%) from set 1 and 13 gaps (3.1%) from set 2 with the former approach. Gaps filled by Sealer had over 90% target correctness (Fig. 4). We think the few outliers with low target percent correctness may be due to low target percent coverage from abnormally large reference gap sequences. These erroneous gap sequences may have been due to differences between the HG38 consensus genome and the NA12878 genome, which resulted in the gap flanking sequences aligning incorrectly.

GAPPadder filled 425 and 411 gaps from sets 1 and 2 (97.9% and 98.8%), but with higher variance on target percent correctness (Fig. 5). The difference in the number of gaps filled, particularly in set 2, and the difference in percent correctness may be explained by GAPPadder using a different gap-filling algorithm than Sealer [6], [11]. However, the overall lower accuracy of set 2 gaps reinforces the notion that gaps in set 2 are more difficult to resolve.

For each gap, we also compared the target percent correctness between the filled gap sequences from Sealer and GapPredict (Fig. 6), and from GAPPadder and GapPredict (Fig. 7). In both Figures 6 and 7, we assigned a target percent correctness of 0% to gaps which Sealer or GapPredict were unable to determine.

We note that for gaps in set 1, there is a cluster at high target percent correctness for both GapPredict and Sealer, and a cluster at low target percent correctness for both tools (Fig. 6). For gaps in set 2, there is a cluster at low target percent correctness for both tools and a cluster at high target percent correctness for GapPredict only.

From figure 7, we note that for both gaps in sets 1 and 2, there is a cluster at high target percent correctness for both GapPredict and GAPPadder, and a cluster at low target percent correctness for both tools. In addition, both figures show outlier gaps where one tool outperformed the other in percent correctness (Fig. 6a, Fig. 7a, Fig. 7b).

Overall, GapPredict predicted 78.9% of the 434 gaps that Sealer originally could fill and 65.6% of the 416 gaps that Sealer originally could not fill with a target percent correctness of at least 90% and a query percent coverage of at least 90%. Generally, we find the gap sequences predicted by GapPredict to be (~1.6 to 1.8x) less accurate compared to those resolved by Sealer and GAPPadder, with the latter two heuristic-based methods yielding sequences having less than 1% base error (Table S1). This is in contrast with the ~1.5% base error on those same predictions generated with the former, machine-learning based method. This is perhaps not unexpected given the GapPredict's paradigm to resolving gap sequences. We expect the sequence accuracy to improve in the future, as machine learning algorithms and models improve.

### 4.3 Model Performance Optimization

With GapPredict, we demonstrate that a deep learning approach shows promise for *de novo* assembly gap predictions. Our analyses showed that GapPredict could predict at least 60% of gap sequences with over 90% target percent correctness and query percent coverage.

Most of GapPredict's model hyperparameters, such as batch size and embedding vector length, were chosen arbitrarily. The number of LSTM cells was explored, as well as the minimum and maximum training k-mer lengths by comparing validation loss and validation accuracy between models trained with different values of these hyperparameters (data not shown). The GapPredict model architecture was based off a simple character-level language model [32]. Further tuning of hyperparameters and exploration of model architectures may produce a better performing model.

Lastly, within each training iteration (but not among every training iteration), the sequences GapPredict trains on are of uniform length. It is possible that allowing these sequences to have variable length could improve training, provided that the length is within the minimum and maximum length hyperparameters. This would prevent each iteration from being biased to a specific input length.

### 4.4 Model Scalability

We trained GapPredict models and filled gaps using these models on a system with two Xeon Silver 4116's, 256 GB RAM, and eight NVIDIA 1080 Ti's. GapPredict models take approximately three minutes to predict a gap sequence using beam search with a beam width of 64. In comparison, it takes approximately 50 minutes to train a model on a gap with about 500 reads mapping to its flanks. Although we can parallelize model training by training different models on different GPUs, the lengthy model training time makes it difficult for GapPredict to scale to the large number of gaps typically present in the draft assembly of larger genomes.

In order to improve runtime, we could employ a stricter patience. With our current patience of 200 epochs, model training often extends past 500 epochs for only a slight improvement in validation loss. We think that the robustness of beam search with a larger beam width can compensate for shorter training. In addition, we could begin training using weights from a pre-trained GapPredict model [33].

One final improvement to our model may be to redesign it to train on the entire read set. Such a model could consider any gap sequence from the source assembly as it encodes data from the entire assembly. This would improve model reusability as a single model could be used to predict any set of gaps for the assembly in parallel, similar in idea to HELEN [17]. In addition, this design would remove the need for obtaining read data for each gap individually, which takes several hours.

### 5 CONCLUSION

With GapPredict, we demonstrate that deep learning is applicable to the *de novo* genome assembly gap-filling problem. Character-level language models indeed seem capable of encoding the information of a gap sequence and its flanks solely by training models on sequence short read data, and use contextual sequencing information for predictions. Further, when such models manage to predict sequence from the reverse DNA strand, they tend to predict the gap with good accuracy. This provides simple ways of filtering out potentially low-quality gap predictions.

Although GapPredict may scale poorly to the high volume of gaps in assemblies for large genomes (>100k gaps for >3Gbp genomes [6]), we think further improvements to speed it up and improve its prediction performance are still possible. In addition, GapPredict was able to provide accurate output (with respect to HG38) for both gaps that Sealer or GAPPadder could fill and gaps that those tools could not fill well or could not fill at all. Deep learning may therefore serve as a method to fill gaps that heuristic methods are unable to fill, rather than being employed as the first and foremost gap closing method. This may lessen the burden of running deep learning tools by decreasing the number of gaps that need to be predicted. In the future, deep learning gap prediction algorithms may complement the current arsenal of gap prediction utilities.

### ACKNOWLEDGMENT

This work was supported by the the National Institutes of Health [2R01HG007182-04A1]. The content of this paper is solely the responsibility of the authors, and does not necessarily represent the official views of the National Institutes of Health or other funding organizations. The authors wish to thank Zhuyi Xue and the Birol lab for their support and ideas during GapPredict's development.

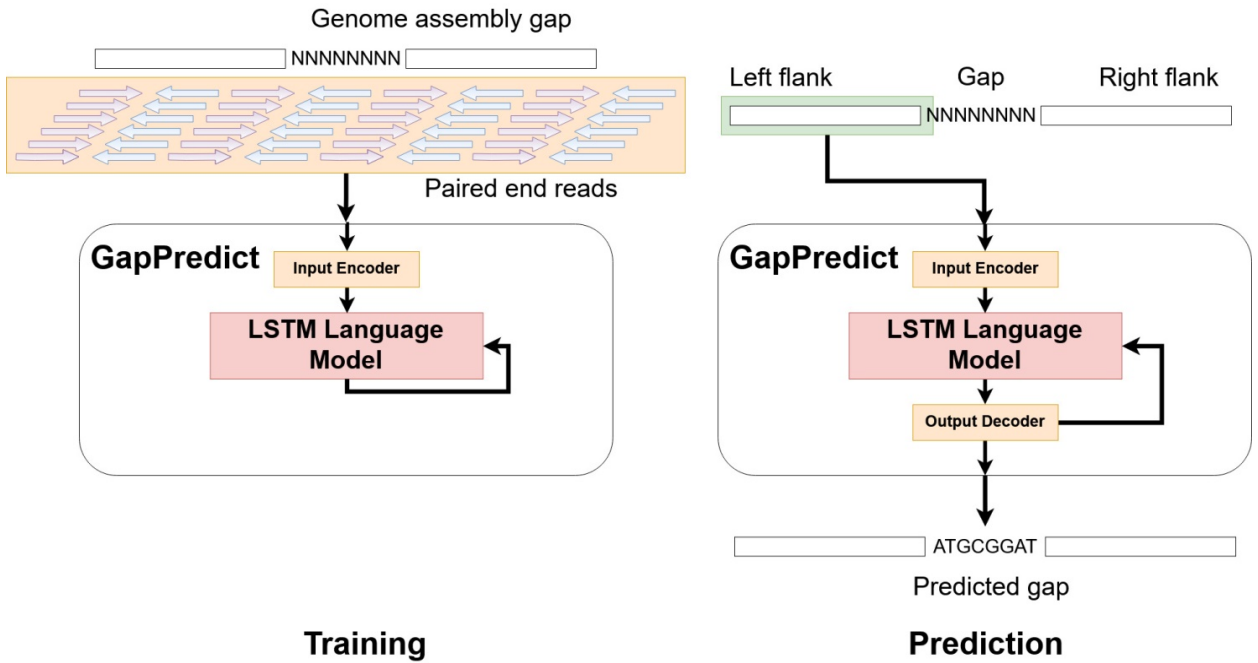

**Fig. 1. Overview of the GapPredict training and prediction process.** During training (left panel), reads mapping to the flanks of a gap are used to train a model capable of predicting the next base given an arbitrarily long input sequence. During prediction (right panel), one of the flanks of a gap (in this case, the left flank) is fed into the model to predict the first base of the gap, providing more context for the model to predict subsequent bases. Notice that if the gap is sufficiently small, paired-end reads aligning to the flanks should cover the entire gap when the coverage depth is high enough.

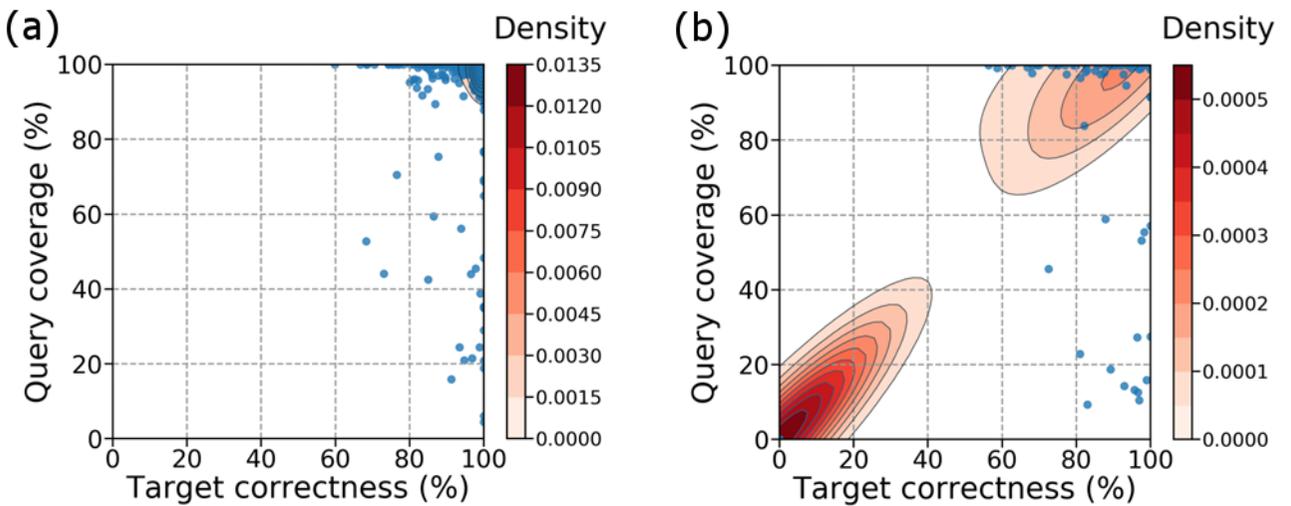

**Fig. 2. Comparison between query percent coverage and target percent correctness for gaps which Sealer was able to fill (set 1). (a) Passing predictions. (b) Failing predictions.** Overall, 78.7% of 868 predictions passed. Colour bars show the density at each level of the contour plot. Kernel density estimation was plotted using default parameters. The figures were generated using Seaborn (v0.9.0).

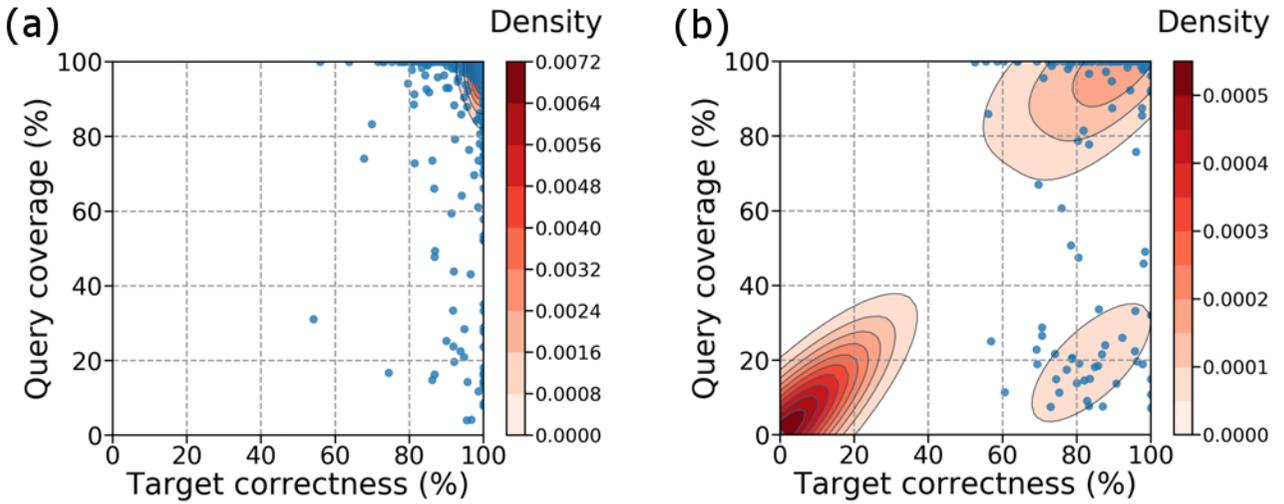

**Fig. 3. Comparison between query percent coverage and target percent correctness for gaps which Sealer was not able to fill (set 2). (a) Passing predictions. (b) Failing predictions.** Overall, 65.2% of 832 predictions passed. Colour bars show the density at each level of the contour plot. Kernel density estimation was plotted using default parameters.

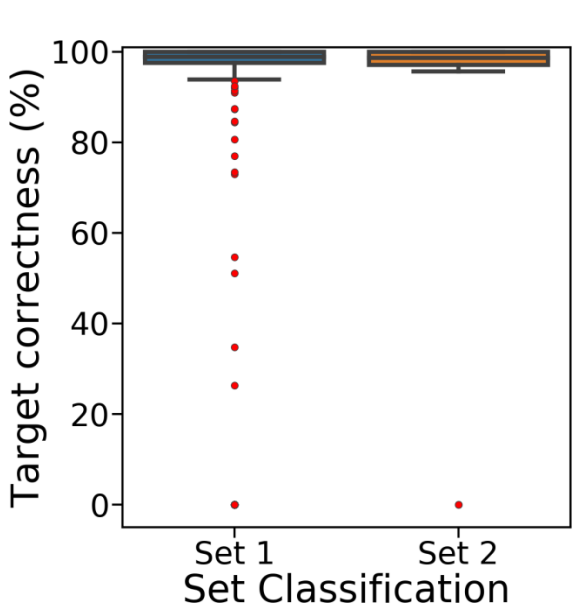

**Fig. 4. Target percent correctness for gaps in set 1 (filled, n=430) and set 2 (unfilled, n=13) that were filled by Sealer when run on each individual gap using only read pairs anchored to the flanks.** From either set, gaps that Sealer did not fill are excluded from the figure.

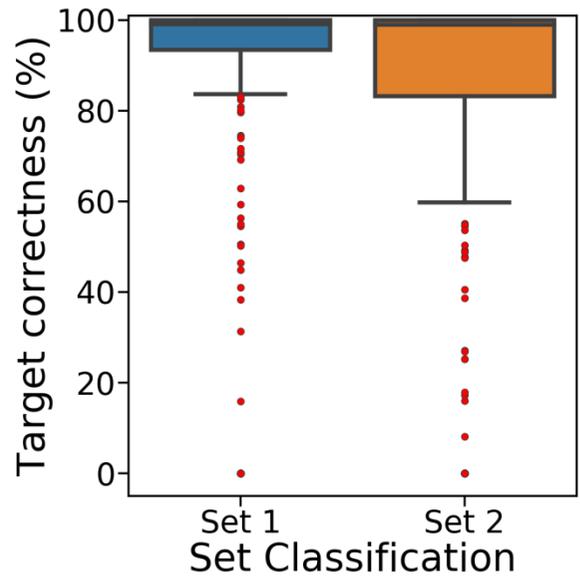

**Fig. 5. Target percent correctness for gaps in set 1 (filled, n=425) and set 2 (unfilled, n=411) that were filled by GAPPadder when run using the full NA12878 draft assembly and all reads in the NA12878 dataset.** From either set, gaps that GAPPadder did not fill are excluded from the figure.

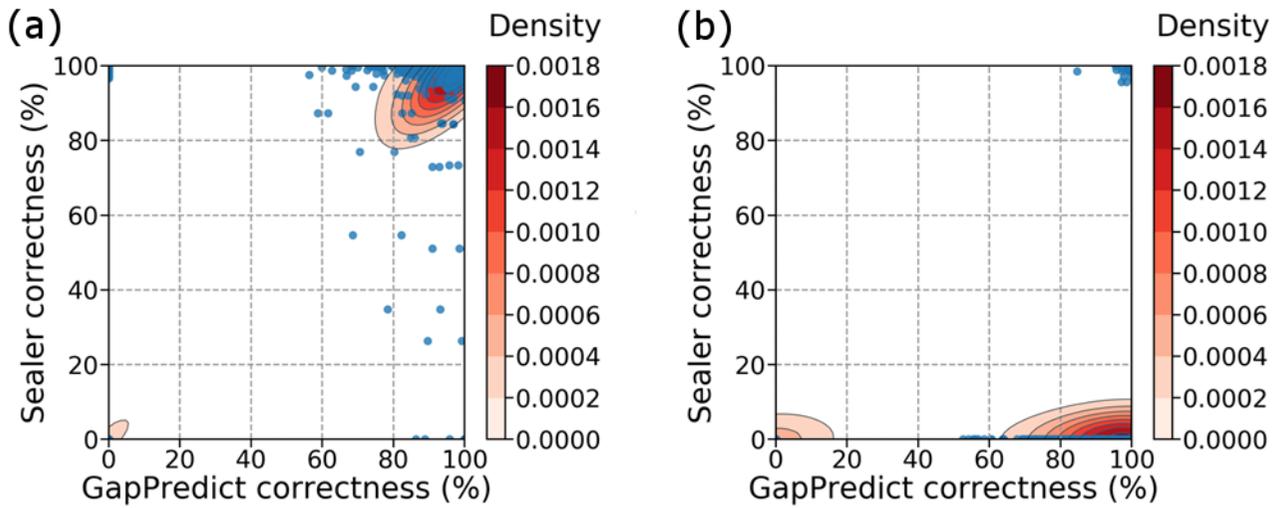

**Fig. 6. Comparison between percent correctness for gaps filled by GapPredict and percent correctness for gaps that Sealer filled when run on each individual gap. (a) Percent correctness for gaps in set 1. (b) Percent correctness for gaps in set 2.** Gaps that Sealer did not fill were assigned a percent correctness of 0%. Colour bars show the density at each level of the contour plot. Kernel density estimation was plotted using default parameters.

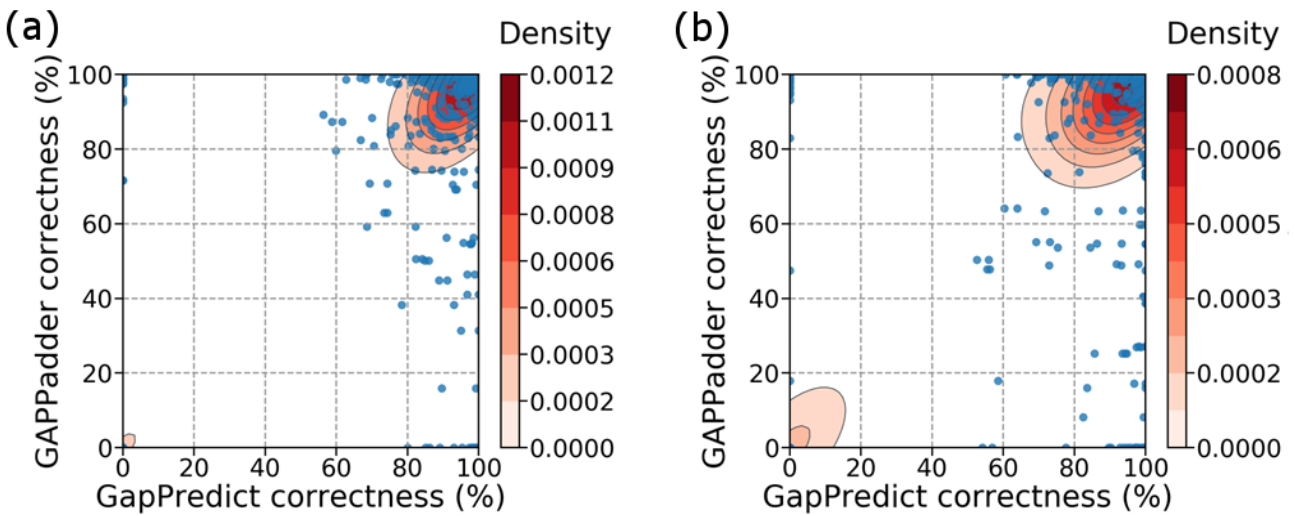

**Fig. 7. Comparison between percent correctness for gaps filled by GapPredict and percent correctness for gaps filled by GAPPadder. (a) Percent correctness for gaps in set 1. (b) Percent correctness for gaps in set 2.** Gaps that GAPPadder did not fill were assigned a percent correctness of 0%. Colour bars show the density at each level of the contour plot. Kernel density estimation was plotted using default parameters.